# A bioinspired optically transparent tough glass composite

*Ali Amini, Adele Khavari, Clayton Molter, Allen J. Ehrlicher\**

A. A., Dr. A. K., C. M., Prof. A. J. E.
Bio-Active Materials Laboratory
Department of Bioengineering, McGill University
Montreal, Quebec, H3A 0C3, Canada
E-mail: Allen.Ehrlicher@mcgill.ca

A. A., Prof. A. J. E.
Department of Mechanical Engineering, McGill University
Montreal, Quebec, H3A 0C3, Canada

**Abstract**

Glasses have numerous applications due to their exceptional transparency, however, poor fracture and impact resistance limit their applications as an engineering material. One relatively recent approach to improve the mechanical properties of materials is through bio-inspiration. Structural biological composites such as nacre, the protective inner layer of mollusk shells, offer far superior mechanical properties relative to their constituents. This has motivated researchers to mimic the design principles in natural composites to create tough transparent materials. However, current bio-inspired materials lack fabrication scalability or offer poor optical transmission. Here, an efficient, scalable bulk process is developed for creating optically transparent tough composites, resulting in a nacreous glass composite material with a four-fold increase in fracture toughness and a three-fold increase in flexural strength compared to conventional structural glasses, and with a 73% of average optical transmittance. The composite consists of glass flakes and poly (methyl methacrylate) (PMMA) assembled utilizing a centrifuge-based fabrication method that aligns and compacts the flakes into layers. To optimize the transparency of the structure, the refractive indices of the PMMA and glass are matched. Based on the results, this nacreous glass composite is proposed as a potential alternative in diverse architectural, vehicular, and electronics applications.

Glasses are generally brittle with low fracture toughness and resistance against impact, limiting their mechanical applications. Glass thermal tempering is a common strategy to increase the strength of glasses,[1] however, this does not dramatically improve fracture toughness.[2] Laminating glass creates a polymeric glass sandwich-like composite structure,[3] the greatest advantage of which is safety: upon fracture, the polymeric layers prevent small pieces of the fractured glass from shattering in catastrophic failure. However, there are only modest mechanical improvements in laminated glasses. [4-6]

To improve glass toughness and impact resistance, researchers have explored bioinspiration - implementing design principles observed in biology. Nacre, the tough material comprising the inner layer of mollusk shells, is a classic example of a tough structural biomaterial; nacre is 3000 times tougher than the components,[7] breaks at 1% of strain – a remarkable improvement relative to the individual ceramic building blocks, and its elastic modulus is approximately 1000 times larger than that of the connective proteins alone.[8]

Many techniques of varying of complexity have been proposed to fabricate synthetic materials mimicking nacre.[9-10] Some of these have focused on making transparent composites,[11-13] resulting in thin films with enhanced mechanical and optical properties. To extend the applications beyond thin films, a new scalable nacreous composite was developed by infiltrating PMMA into a glass flake scaffold while matching refractive indices of the two phases.[14] Despite superior fracture resistance properties compared to glass, this composite sacrificed transparency, a key feature for widespread applications.

In contrast, others have employed top-down methods, including laser-engraving interlocking jigsaw-shaped 3D arrays in bulk glass,[15] and glass lamination processes of thin glasses with laser-engraved cross-plied[16] and tablet-like architectures.[17] These approaches resulted in increased composite fracture toughness and impact resistance, but reduced stiffness and strengths. Stiffness and strength can be generally improved by decreasing the size of the patterns, however, this reduces transparency and scalability.[17] This highlights the general trade off challenge that bio-inspired glasses have suffered from between mechanics, transparency, and fabrication scalability. While these diverse strategies have explored bottom-up and top-down approaches resulting in excellent mechanical, optical, and fabrication results, no method has successfully combined all three together in a tough glass.

In this paper, we demonstrate a bottom-up fabrication technique to produce a transparent brick and mortar structural composite possessing advantageous mechanical properties that improve upon those of normal glasses or their bio-inspired composite counterparts, as shown in **Figure 1**. Here for our composite, we use glass flakes and Poly (methyl methacrylate) (PMMA) for the hard component and soft phase, respectively. Glass flakes were utilized as the hard component due to their high diameter-to-thickness aspect ratio, transparency, large elastic modulus, and well-characterized surface chemistry for surface functionalization. PMMA, an amorphous polymer that is polymerized through a free radical bulk polymerization process,[18] was selected as the soft phase due to its high strength and stiffness, relatively large yield strain,[19] and excellent optical properties.[20] We functionalized the glass tablet surface with a silane to strengthen the bond between soft and hard phases. To make the nearly opaque nacreous composite transparent, we increased the refractive index (RI) of PMMA to that of the flakes by adding an organic dopant, phenanthrene, to our polymeric matrix.[21] We then imposed an aligned brick and mortar architecture and high volume-fraction by centrifuging the composite, and finalized PMMA polymerization by baking.

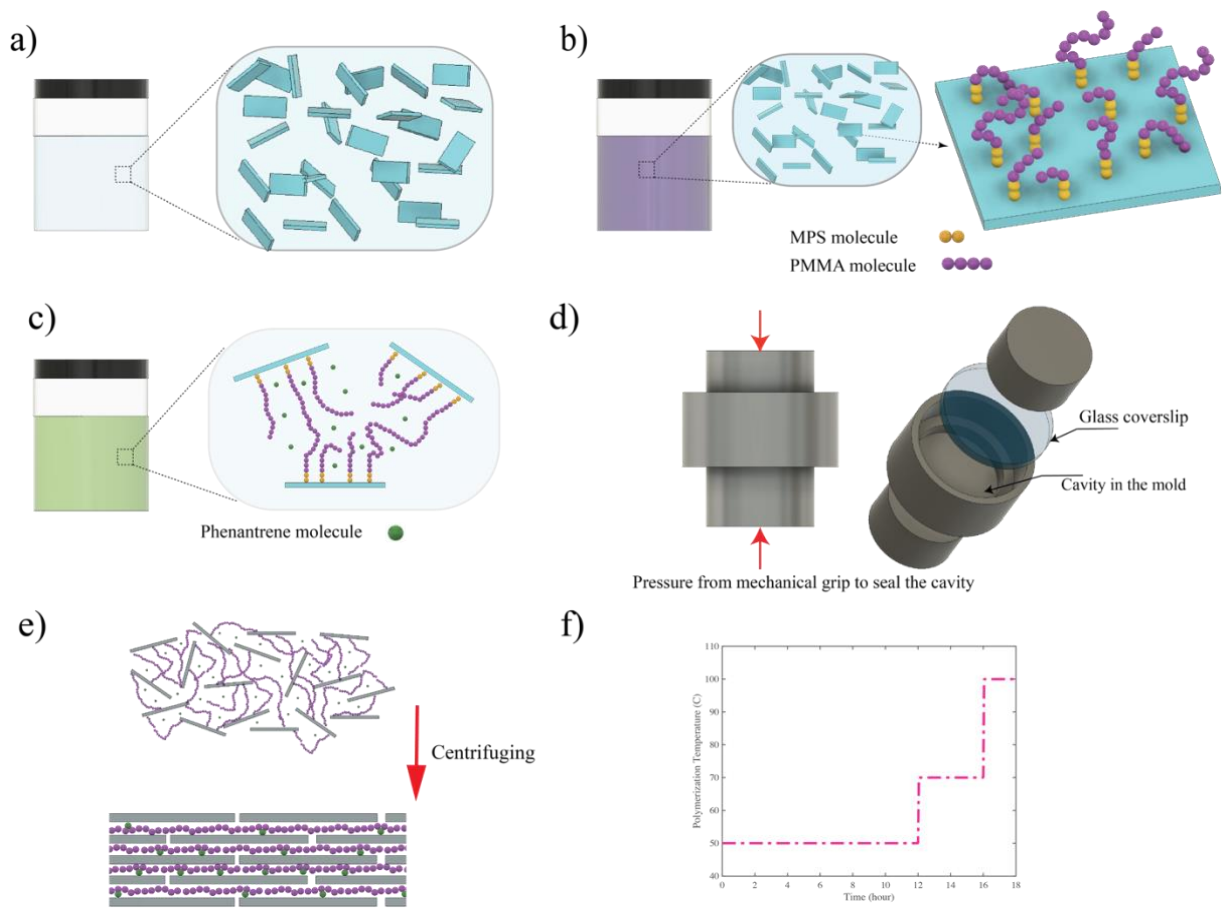

**Figure 1.** Centrifuged-based fabrication method of nacreous glass composite. a) Cleaned glass flakes were dispersed in toluene. b) Glass flakes surface-treated with ɣ-MPS and then with a solution of MMA in toluene to promote polymerization from the glass surface. c) Surface-treated glass flakes were involved in free radical polymerization of PMMA at 50°C. d) Glass-PMMA mixture transferred to a casting mold with a pre-designed cavity on the bottom e) Glass-PMMA mixture was centrifuged to impose alignment in flakes and densify the mixture. f) Polymerization process finalized in oven: 12 hours at 50°C, 4 hours at 70°C and 2 hours at 100°C.

We found the optimum dopant concentration to be 12%, yielding an average spectral transmittance of 76% for a 1 mm glass composites (**Figure 2-a** and **Figure 1S-b**). Our glass composite's transmittance compares well with both soda-lime monolithic glass and PMMA doped with 12% of phenanthrene (Figure 2-b and 2-d). It also has 24% higher transmittance than similar bio-inspired laminated composites,[17] and has almost 100% higher transmittance than nacre mimetic bulk-fabricated composites.[14] While our composite is hazier than the soda-lime glass with the same thickness(Figure 2-c and 1S-c), it is more than 70% less hazy than similar bio-inspired bulk fabricated composites.[14]

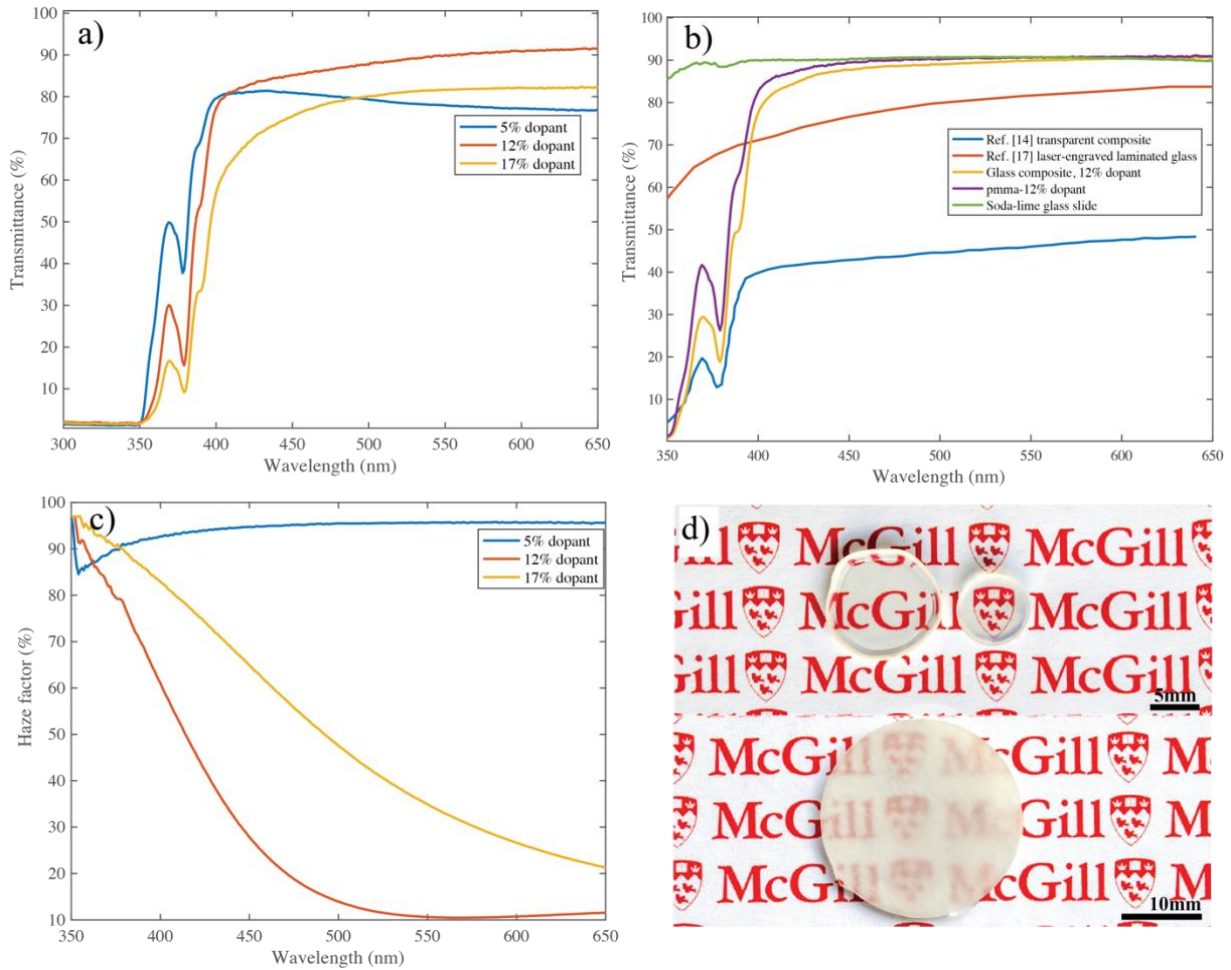

**Figure 2.** Dopant percentage and composite thickness affect composite transparency and haziness. a) Transmittance values for 1mm thick samples and different phenanthrene weight percentages. b) Transmittance for the 12% glass composite, doped PMMA (12%), soda-lime glass, bio-inspired transparent composite,[14] and laser-engraved laminated glass.[17] Our composite compares well with soda-lime monolithic glass and is superior to its bio-inspired counterparts. c) Haze factor values for 1mm thick composites and different dopant weight percentages. d) 1mm thick glass composites with 12% dopant (top), and 0% dopant (bottom).

As glass and PMMA have different densities, we used centrifugation to increase the fraction of glass in our composite, leading to a high volume fraction of the stiff (glass) phase and consequently a thin connective (PMMA) phase (**Figure 3-a**). Centrifugation also yielded a structure with more aligned flake orientation (Figure 3-b and 3-c).

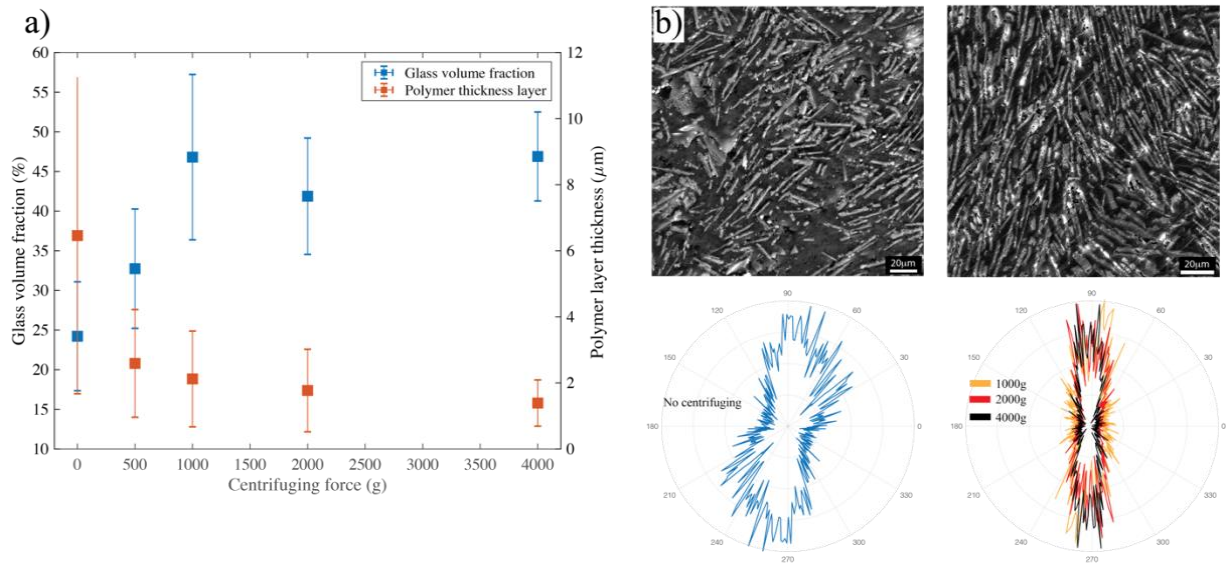

**Figure 3.** Centrifugation increases the glass volume fraction by decreasing the polymer thickness layer between tablets. a) Glass volume fraction increases almost two-fold when the sample centrifuged (2000g). Also, polymer layer thickness decreases about 50% when centrifuged with 2000g force. b) Section SEM image of a non-centrifuged composite (top, about 24% glass volume fraction). We observed a noticeable number of areas with no flakes, and also many flakes with random orientation in the material. Polar distribution of orientation in the flakes also confirms this observation (bottom). c) Section SEM image of centrifuged composite (2000g, about 43% glass volume fraction). Flakes are more oriented in one direction, and areas with no flakes are rarely observed. Polar distribution of orientation in the flakes for different centrifuging speeds also shows insignificant difference between 1000g, 2000g and 4000g forces. Data points and error bars are mean values and standard deviation respectively.

Using the 3-point bending test, we found that surface functionalization with γ-MPS increased the final strength for the glass composite two-fold (**Figure 2S-a**). This increase in strength, however, produced a composite only slightly stronger than pure PMMA (**Table 1S**). The final strength was increased to about 140 MPa by including the centrifuging process as a part of fabrication process; this aligned the glass flakes into layers of parallel planes and also yielded a denser overall structure. The beneficial strengthening effects of centrifugation appeared to plateau at 2000g, with no significant increase in flexural strength for higher forces (**Figure 4-a**). The flexural modulus was also increased from 4.7GPa for non-centrifuged sample to about 7.2Gpa for the sample centrifuged with 2000g force, with no significant increase of the modulus with higher centrifugation speeds. The rupture strain decreased about 17% for the surface-functionalized samples compared to the non-functionalized one. This suggests that the strong bond between the soft and the hard phases in the functionalized composite limits the displacement in the polymeric phase; however, we observed an increase in the rupture strain for the centrifuged samples (Figure 3S-b).

Single-Edge Notch Bending (SENB) tests were performed to evaluate the fracture resistance of our glass composites by calculating the crack initiation fracture toughness ($K_{IC}$) and the work of fracture (*WOF*).  While the non-centrifuged composite sample has a higher fracture strength compared to the pure PMMA, their fracture stress-strain curves are very similar (Fig 2S-c). In other words, despite the smoother crack propagation in the glass composite relative to the pure PMMA, the fracture is still catastrophic in both cases. The centrifuged glass composite, on the other hand, lacked catastrophic fracture and displayed crack propagation with higher fracture strength. Centrifuging (2000g) increased the $K_{IC}$ from 1.75 MPa m$_{0.5}$ for the non-centrifuged samples to about 2.25 MPa m$_{0.5}$ (Figure 4-b). Also, by centrifuging (2000g), we increased the *WOF* for the composite from 308 J m$_{-2}$ to 405 J m$_{-2}$. The results in Figure 4-b implies that the centrifuging process promotes the toughening mechanisms of tablet sliding coupled with polymer phase stretching and tearing (Figure 4-c), and tablet pull-out (Figure 4-d). Due to the activation of such toughening mechanisms, and that the crack grows mainly through the soft phase and displaces tablets, many deflections in the crack propagation path are observed in the macro-scale (Figure 4-e). Our composite compares favorably with current state of the art materials in terms of fracture toughness and strength (Figure 4-f). Considering the *WOF* as a non-linear measure of fracture resistance, our composite outperforms annealed and laminated glasses and also pure PMMA (Figure 4-g). The laser-engraved laminated glass[16] possesses a very high *WOF*, however, this has been achieved only with a reduction in strength.

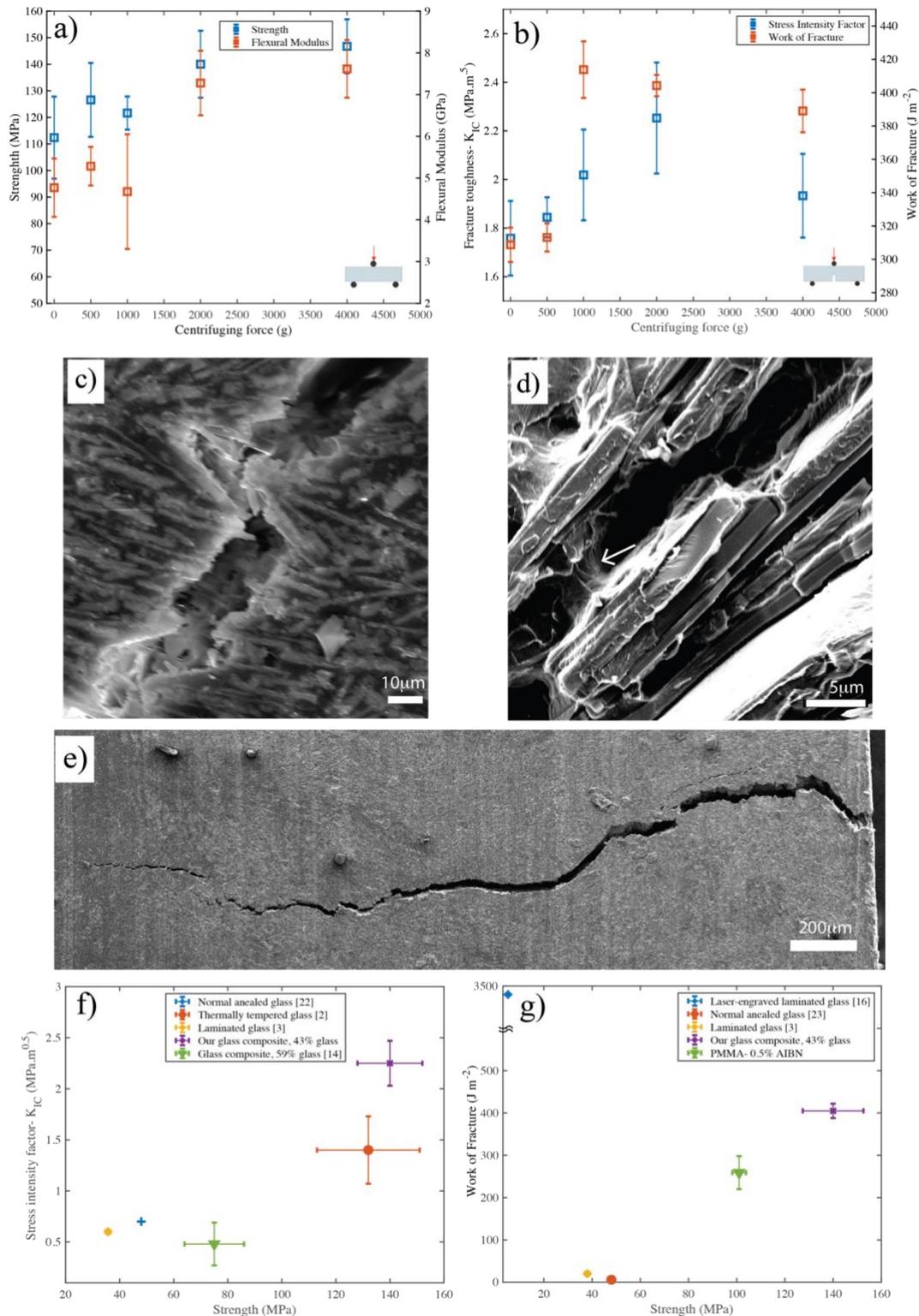

**Figure 4.** Nacreous glass composite outperforms normal, tempered and laminated glass under fracture. a) strength and flexural modulus increase with centrifugation. While the higher centrifuging speed yields higher strength and flexural modulus, 2000g appears to be the saturation point. b) K$_{IC}$ and *WOF* values increase with increasing the centrifuging speed up to 2000g. c) Tablet pull-out is one of the most important extrinsic toughening mechanisms in microscale. d) Tablet sliding causes polymer stretching and tearing, and in large deformations, polymer bridging between tablets (white arrow). e) Crack deflection in the material as a result of microscopic toughening mechanisms. f) Crack initiation fracture toughness vs final strength for normal annealed soda-lime,[22] tempered,[2] laminated,[3] bio-inspired transparent

composite,[14] and our composite glass (2000g). The composite demonstrated here outperforms other state of the art glasses in both fracture toughness and final strength. g) *WOF* versus final strength for normal annealed,[23] laminated,[3] and laser-engraved laminated [16] glasses, as well as the pure PMMA and our composite glass(2000g). The only glass that exceeds our composite in terms of *WOF* is [16], but it possesses far less strength compared to our composite. Data points and error bars demonstrate the mean value and the standard deviation respectively.

Here we demonstrate a novel nacreous composite structure with previously unattained optical and mechanical properties that could be a potential alternative to monolithic and laminated glasses. We combined glass flakes and PMMA as the hard and soft components of our composite respectively and used phenanthrene as a dopant to increase the PMMA's refractive index and match it to the glass flake's, resulting in a remarkably strong yet optically transparent material. Furthermore, to recreate the highly aligned hard tablet phase found in nacre, we centrifuged the glass-PMMA mixture, dramatically decreasing the inter-tablet spacing and increasing the glass volume fraction, thus creating a dense multilayered stack of aligned tablets bound together by thin PMMA films. This structural compaction and alignment significantly increased the flexural strength and fracture toughness of the composites, as it imposes order on the structure and paves the road for activation of toughening mechanisms such as tablet pull-out, crack deflection and tablet sliding. In the absence of mineral bridges and tablet interlocking, tablet sliding is likely the most important mechanism responsible for high fracture toughness in our composite. Large deformations in polymeric phase in forms of stretching and tearing cause yielding, and consequently, plastic deformation in the PMMA (non-linear part in figure 2S-a). This leads to large deformations and high levels of energy absorption, as well as potential polymeric bridges between tablets that consequently resist their sliding and increase the fracture toughness. Indeed, tablet sliding is key to the exceptionally high *WOF* in [16] and appears to be absent in [14]; this is likely the reason why our current composite outperforms the material in [14] in terms of fracture toughness. Our glass composite also outperforms annealed, thermally tempered, and laminated glasses in fracture toughness and final strength. Finally, our composite shows an excellent level of transparency, with only 16% less than the one for the monolithic glass (for 1mm of thickness).

We hope that the composite described here may find applications in scalable performance transparent composites. Moreover, the strategies for fabrication presented here allow researchers to tune the final composite mechanical and optical properties, as well as apply them to other hard phase tablets.

**Experimental Section**

*Glass surface functionalization*: Glass flakes were cleaned in the Piranha solution, washed in DI water, dried in a vacuum oven, and then functionalized in a solution of toluene and the silane agent (ɣ-MPS) for 12 hours. The surface-treated glasses were washed afterward and again dried in a vacuum oven. To promote PMMA polymerization from the glass surface, thin PMMA monolayers were grown on the glass surface by involving the flakes in a free radical polymerization process, in the presence of excess toluene.

*Glass composite fabrication:* To match the refractive indices of the glass and PMMA, we dissolved phenanthrene in MMA. The phenanthrene-doped MMA, the initiator, and the glass flakes were mechanically stirred at 50$_o$C until a more viscous mixture was obtained. The mixture then was cooled immediately and transferred to a 3D-printed polypropylene casting mold. To make a dense structure with well-aligned flakes, we centrifuged the glass-PMMA compound in a two-step process: a low-speed centrifugation step to induce alignment in the flakes, and a final high-speed centrifugation step to densify the structure. The casting mold was sealed, and the composite was baked in the oven.

*Estimation of glass volume fraction, polymer layer thickness and orientation distribution:* We calculated the glass volume fraction by Archimedes' principle, assuming that the composite only consists of the hard and soft phases. The volume of the composite samples measured using a pycnometer (VWR, 25 ml, Gay-Lussac type), and by knowing the densities of the glass flakes and the PMMA, we estimated the glass volume fraction in our composite.

The polymer layer thickness was measured by extracting data from several line scans on the SEM images of the composite cross-sections in ImageJ software.

The orientation distribution was measured and plotted by analyzing the SEM images of the composite cross-section using a MATLAB code based on the [24].

*Structural characterization of the composite (SEM images):* Samples were imaged using a scanning electron microscope (FEI Quanta FEG 450) at (20KV at secondary electron mode) to evaluate the ordering of the flakes with respect to the centrifugation speed. The samples were initially coated with a layer of platinum (4nm) using a sputter coating machine (Leica Microsystems EM ACE600 High Resolution Sputter Coater).

*Optical characterization of the composites:* PMMA samples doped with phenanthrene were dissolved in toluene and then coated on silicon wafers. Refractive index of the samples then measured using a spectroscopic ellipsometer (Sopra GES-5E). The transmittance of the glass composites measured using a UV-Vis-NIR spectrophotometer (LAMBDA 750 UV/Vis/NIR). For this purpose, cylindrical glass composite samples with the various diameters and thicknesses were prepared.

*Mechanical characterization of the composite*: To measure the elastic modulus, final strength, and rupture strain of the composites, 3-point bending tests were performed using a universal testing machine (Admet, eXpert 5000, MA US). Cubic samples with dimensions of 25x3.2x1.8 mm were prepared based on standard ASTM D790.[25] Support span and displacement rate was 16 mm and 1 um/sec respectively.

Fracture toughness of the composites was evaluated using Single-Edge Notched Beam (SENB) test and according to the ASTM E1820 standard.[26] Cubic samples with 25x3.2x1.8mm dimensions were prepared, and a notch was created using a 450 µm diamond saw. The initial crack (40 µm in tip radius) then created on the tip of the notch using a thin blade covered with diamond paste. The samples then were used in a 3-point bending set up with a displacement rate of 1 um/s.

**Acknowledgements**
AJE acknowledges support from NSERC RGPIN/05843-2014 & EQPEQ/472339-2015, FRQNT team grant, Canadian Foundation for Innovation Project #32749, and the Canada Research Chairs Program. Authors' contributions: Conceptualization, A.J.E., A.A.; Methodology- development, A.A., A.K., A.J.E.; Methodology-application, A.A.; Investigation, A.A., A.K.; Formal Analysis, A.A.; Software, A.A.; Visualization, A.A.; Writing- original draft, A.A.; Writing- review and editing, A.J.E., C.M., A.K., A.A.; Funding acquisition, A.J.E.; Resources, A.J.E; Supervision, A.J.E.

# Supporting Information

*Materials*

Glass flakes (GF001-10, d50 (median particle) diameter =27-32 μm, thickness =0.9-1.3 μm, refractive index = 1.52) were kindly supplied by Glassflake Ltd. Methyl methacrylate (MMA, 99%), azobis isobutyronitrile (AIBN, 98%), Phenanthrene (98%), (3-trimethoxysilyl) propyl methacrylate (γ-MPS, 98%), Methanol (ACS, 99%), hydrogen peroxide (30 wt. % in H2O), acetone (99.5%) and MMA inhibitor remover were acquired from Sigma . Toluene (reagent grade) and sulfuric acid (reagent grade) were purchased from Caledon Laboratories Ltd.

*Glass surface treatment*

Glass flakes were cleaned in Piranha solution (3 parts of concentrated sulfuric acid, 1 part of water and 1 part of 30 wt.% hydrogen peroxide solution) for 30 minutes, and subsequently washed in DI water several times and dried in a vacuum oven at 120°C overnight. Cleaned and dried flakes (2 g) were mixed with a solution of toluene (15 ml) and the surface functionalization agent (5 ml), γ-MPS, (3:1 volume ratio) and gently stirred with a magnetic stirrer for 12 hours. The silane functional groups will react with hydroxyl groups on the glass surface. The surface-treated glasses were washed with toluene, methanol, and then washed with DI water several times and dried in a vacuum oven at 110°C for 2 hours. γ-MPS functionalized glass flakes were then involved in a free radical polymerization process to grow PMMA monolayer on their surface. This was performed in a two-step process: first, the glass flakes were added to a mixture of dry toluene and MMA (2:1 volume ratio) under gentle mechanical stirring at 70°C for 30 minutes with AIBN (1 wt.%) as initiator. This is to promote growing PMMA from the glass surface and decelerate the polymerization in bulk MMA. After this step, we dried the flakes in a vacuum oven at 110°C for 2 hours prior to usage in the main composite fabrication process, which follows in the next section.

*Fabrication process*

To adjust the refractive indices of the glass and PMMA, we dissolved an aromatic hydrocarbon, phenanthrene, in MMA as a dopant. Phenanthrene and AIBN (0.5 wt.%) were dissolved in MMA and added to the surface-treated glass flakes. The mixture was mechanically stirred at low speed for 45 minutes at 50₀C under argon atmosphere and immediately cooled down in an ice-water bath afterward. The glass-polymer mixture then transferred to a 3D-printed polypropylene casting mold containing a cavity with desired shape and depth. The cavity's depth will be the glass composite's final thickness. To make a dense structure with well-aligned flakes, we centrifuged the glass-MMA compound. The centrifuging process involved two low-speed steps (100g RCF for 5 minutes and 300g RCF for 5 minutes) to induce alignment in the flakes. Then we performed the last step in high speed (more than 1000g RCF for 20 minutes) to make a denser structure. The top of the cavity was covered by a glass coverslip, and a gentle pressure applied from a small mechanical grip sealed the centrifuged glass-PMMA composite in the cavity. We finalized the polymerization process by exposing the composite to heat in the oven (50₀C for 12 hours, 70₀C for 4 hours, and 100₀C for 2 hour).

*Composite centrifugation*

We tested various centrifugation speeds and measured the orientation distribution of the flakes and the resulting volume fractions. Centrifuging appeared to have a significant effect on the orientation of the glass flakes (Figure 3-b and 3-c bottom). By comparing the polar orientation distribution graphs of the non-centrifuged and centrifuged composites, the role of the centrifuging process in inducing order in the composite's structure can be observed. Although flakes in 2000g-centrifuged samples were well oriented compared to the non-centrifuged samples, increasing the centrifuging speed seemed to have a negligible effect on further orienting the flakes in one direction. Regarding the effect of the centrifuging process on the glass volume fraction, we found an effective saturation centrifugation force, after which the increase in composite volume fraction appears negligible (Figure 2-a). Glass volume fraction increased from about 24% for non-centrifuged composite to about 43% for the centrifuged sample (2000g). Since no drastic change in volume fraction was observed for more increasing of the centrifuging speed, and also considering the mechanical testing data, we determined 2000g to be an optimal centrifugation force. The polymer layer thickness between flakes also decreased dramatically by applying centrifuge forces. The polymer layer thickness decreased from about 35μm for the simply mixed sample to about 17μm for the sample centrifuged with 2000g.

*Index matching the composite components*

We realized that by combining two materials with excellent optical properties, we might create a composite which had similar transparency. Such a material would have clear advantages in diverse applications. However, initially preparing these materials, we found they were nearly opaque with a white appearance due to the strong scattering of light at the multiple interfaces between glass and PMMA. We matched the refractive indices of the two phases using phenanthrene as a dopant to improve the transparency. To find the percentage of dopant that optically optimized the overall composite transparency, we first measured the refractive index of PMMA-dopant samples as a function of dopant weight percentage (**Figure 1S-a**). We estimated that between 12 to 16 % of dopant along with 0.5 % AIBN as polymerization initiator would match the RI of PMMA to the one of our glass flakes (1.524). Knowing the refractive indices as a function of PMMA formulation and dopant, we can estimate the composition that leads to the highest level of transparency. To reach the optimum dopant percentage, composites with 1mm of thickness, 43% volume fraction, and various phenanthrene amounts were made. We measured the transmittance of the composites as a function of composition by a UV-vis spectrophotometer (Figure 2-a). The optimal dopant percentage was determined to be 12%, yielding an average transmittance of 76% for our 1 mm glass composites (Figure 1S-b). Our glass composite's transmittance compares well with both soda-lime monolithic glass and PMMA doped with 12% of phenanthrene for wavelengths higher than 400nm (Figure 2-b). However, the transmittance values declined dramatically for the wavelengths smaller than 400nm, again resembling the behavior of 12% phenanthrene-PMMA. The average transmittance for our composite is only 16% less than the soda-lime glass (Figure 1S-c). Despite the high transmittance values for the glass composite, this material also tends to diffuse light resulting in a hazy appearance (Figure 2-c). The material appeared to have the lowest haziness for the optimal amount of dopant (Figure 1S-d). Although even 5% of dopant yielded a reasonably high average transmittance, the sample possessed an unsatisfactorily high haze factor of about 90% for almost the whole visible light spectrum. The haze factor for the composite with 12% of dopant, however, is not constant for the whole visible light spectrum and increases from about 10% for wavelengths greater than 500 nm to about 90% at the low end of the visible spectrum. Although a high level of transparency is a feature in common between index-matched composites (Figure 1S-e), the average transmittance drops 25% as the thickness increases from 1mm to 3mm. The haze factor value also increased as a function of increasing the thickness as it appears in the haze factor curves (Figure 1S-f), and as shown in Figure 1S-g and 1S-h. The composite with no dopant (Figure 3-d, bottom), on the other hand, was very hazy, and the sample was not transparent due to the light diffusion.

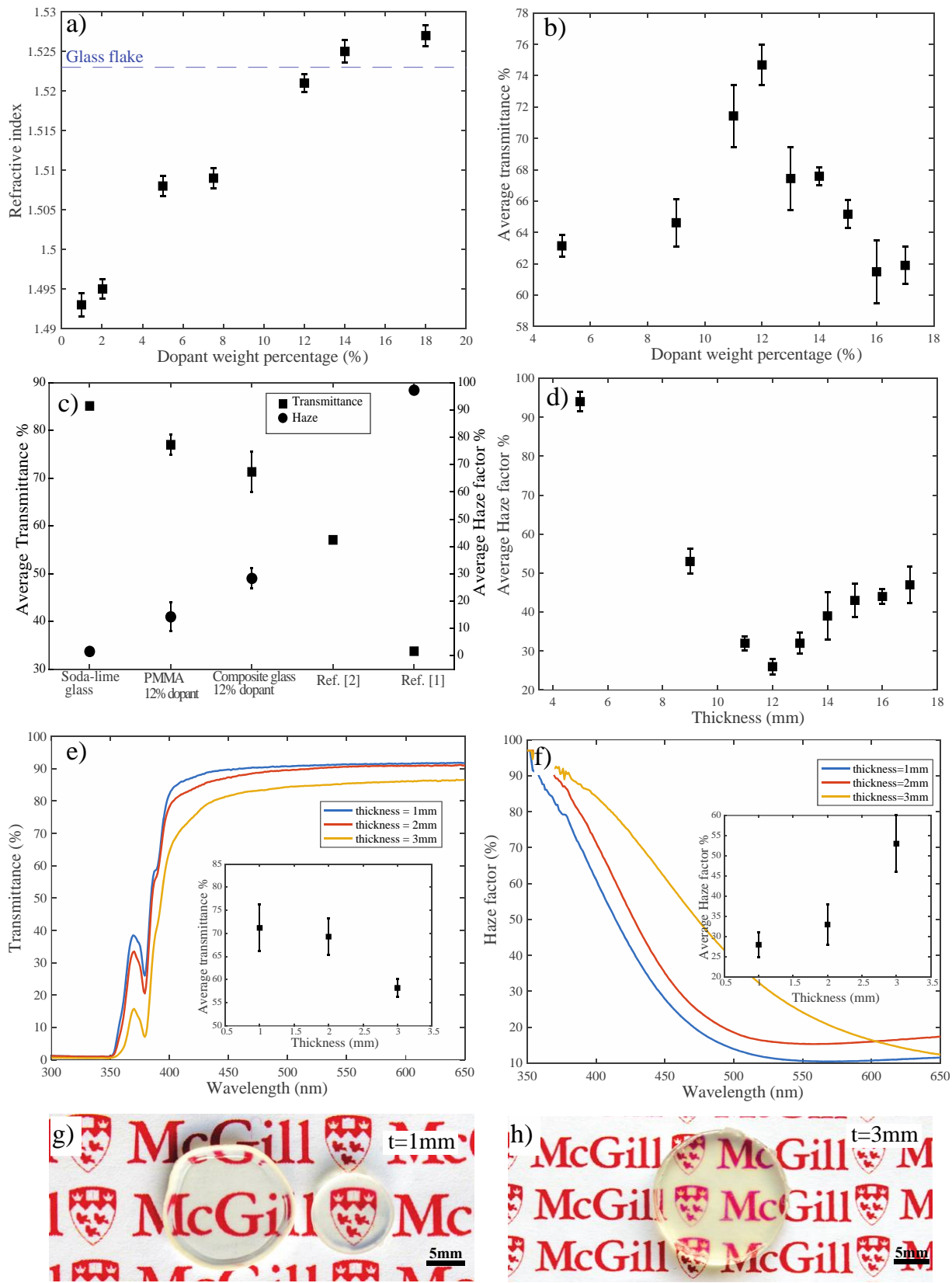

**Figure 1S**. Glass composite doped with Phenanthrene is highly transparent, but hazy. a) PMMA refractive index increases linearly by increasing phenanthrene weight percentage. b) Average transmittance values in terms of dopant weight percentage. 12% appears to be the optimum dopant amount. c) Comparison of transmittance and haze factor for the 12% glass composite, doped PMMA(12%), soda-lime glass, bio-inspired transparent composite,[1] and laser-engraved laminated glass.[2] Although the transmittance of our glass composite is similar to the soda-lime glass, its haze factor is higher by a factor of 20. Our glass composite, however, is superior to its bio-inspired rivals both in transmittance and haze factor. d) Haze factor values for 1mm thick composites and different dopant weight percentages. 12 wt% of dopant yields the lowest haze factor value. e) Increasing the glass composite thickness decreases the average transmittance. A decline of 25% is observed in average transmittance as the thickness increased from 1mm to 3mm (inset). f) Composite thickness increase affects the haze factor negatively. The haze factor increases from 28% to 52% for 1mm and 3mm thicknesses respectively(inset). g) Glass composites with 12% dopant and 1mm of thickness. h) Glass composites with 12% dopant and 3mm of thickness. It looks hazier than the one with 1mm of thickness. Data point and error bars demonstrate the mean value and the standard deviation respectively.

*Performance of the composite under mechanical loading*

A significant increase of final strength observed for the glass composite surface-functionalized with γ-MPS, compared to the one without any surface treatment (**Figure 2S-a**). Unlike the composite with no surface functionalization, which has only one linear regime, the surface-functionalized composite experiences two distinct linear and non-linear regimes during the flexural testing. The non-linear regime is attributed to the large plastic deformation of the PMMA after yielding. The final strength of the composite was increased about two-fold by functionalizing the glass flakes surfaces by γ-MPS. This increase in strength, however, only produced a composite slightly stronger than pure PMMA (**Table 1s**).

**Table 1S.** Strength, rupture strain, and flexural modulus values for composite with no surface functionalization and centrifuging, surface-functionalized and centrifuged, and PMMA samples.

|  | Strength [MPa] | Rupture strain [%] | Flexural modulus [GPa] |
|---|---|---|---|
| No surface functionalization and centrifuging | 56.17 ± 4.30 | 3.37 ± 0.14 | 2.46 ± 0.31 |
| Surface-functionalizing and no centrifuging | 112 ± 15.21 | 2.77 ± 0.88 | 4.77 ± 0.75 |
| Surface-functionalized, centrifuged (2000g) | 140.01 ± 12.62 | 3.05 ± 0.31 | 7.27 ± 0.77 |
| PMMA | 101 ± 2.85 | 6.82 ± 0.39 | 1.41 ± 0.42 |

Increasing the centrifuging speed densifies the composite structure and increases the volume fraction of the hard phase, which consequently leads to a higher flexural modulus. But similar to the final strength values, 2000g seems to be the saturation point and increasing the speed does not impose any significant effect on the flexural modulus values. The effect of increasing the centrifuging speed, however, appeared to be insignificant for the rupture strain, and all the samples possessed a rupture strain of about 3%.

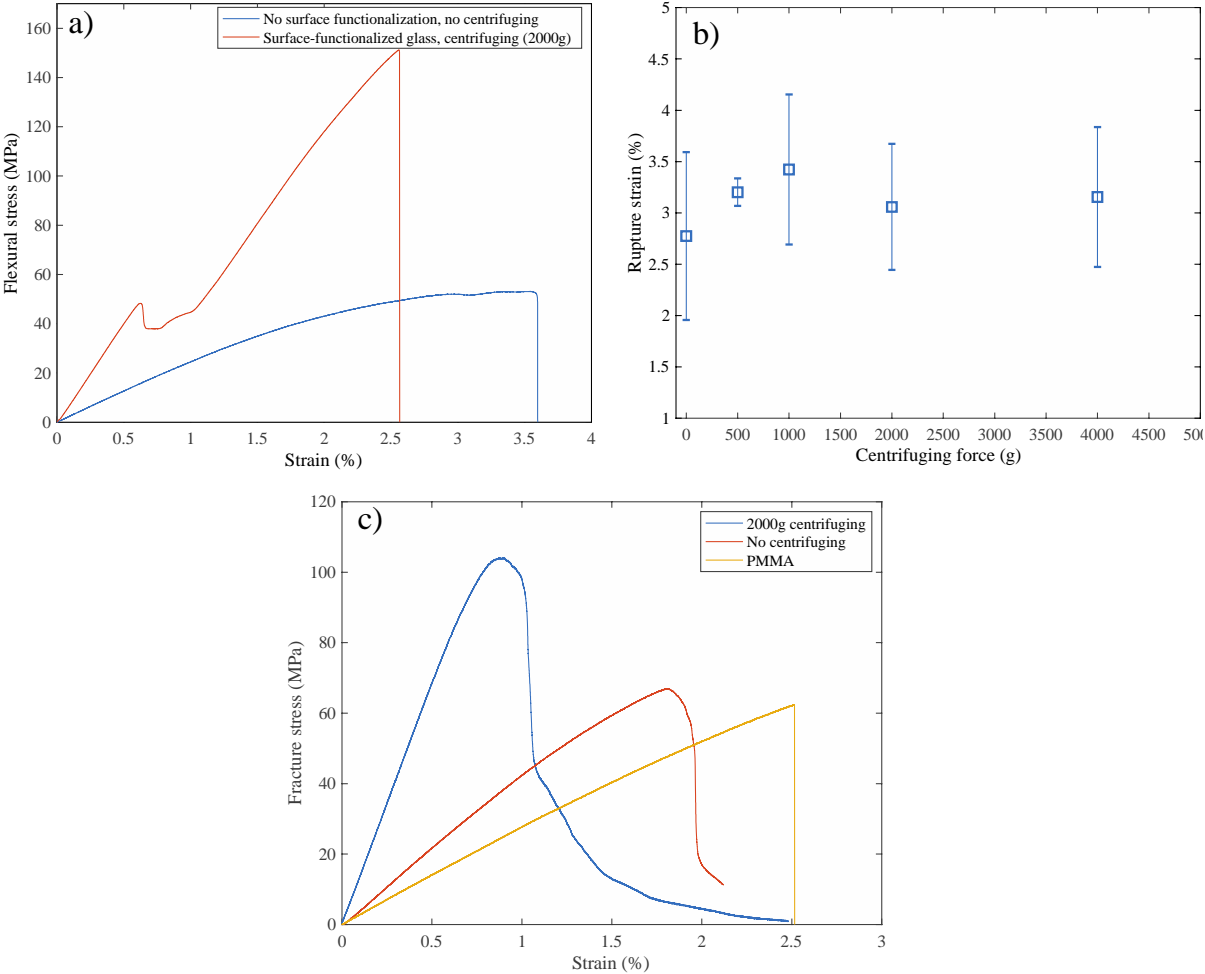

**Figure 2S**. a) Fracture stress-strain curves for the composites with and without surface-functionalized glass. Surface-treated composite experiences two deformation regimes, a linear regime followed by a non-linear one up to the failure, whereas the composite with no glass functionalization deforms linearly to the fracture. b) Effect of centrifugation on rupture strain of the composite. Centrifuging increases the rupture strain about 20% when centrifuged with 500g, but the effect seems to saturate after that. c) Fracture stress-flexural strain curves for pure PMMA, non-centrifuged and centrifuged composites, illustrating an increased fracture strength with composite formulation and subsequent centrifugation.
21

*Fracture mechanics calculations*

To compare the maximum load the notched samples can bear, we defined fracture stress, $\sigma_f$, as a function of the applied load and the un-notched ligament of the sample:

$$\sigma_f = \frac{3f}{B(W-a)^2}$$

The crack initiation fracture toughness, $K_{IC}$, was calculated from the load-displacement curves and based on the maximum force value and initial crack size from the following equation:

$$K_{IC} = \left(\frac{P_c S}{W\sqrt{WB}}\right) f\left(\frac{a_c}{W}\right)$$

where $P_c$ and $a_c$ are maximum amount of load and initial crack size, respectively, and $S$, $W$ and $B$ are support span, specimen width and thickness, respectively. Also,

$$f\left(\frac{a_c}{W}\right) = \frac{3\sqrt{\frac{a_c}{W}}\left(1.99 - \frac{a_c}{W}\left(1 - \frac{a_c}{W}\right)\left(2.15 - 3.93\left(\frac{a_c}{W}\right) + 2.7\left(\frac{a_c}{W}\right)^2\right)\right)}{2\left(1 + 2\frac{a_c}{W}\right)\left(1 - \frac{a_c}{W}\right)^{\frac{3}{2}}}.$$

The work of fracture (WOF) was calculated as a nonlinear measure of fracture toughness. WOF is defined as the total energy spent to create one unit of fracture surface area[3] and calculated as follow

$$WOF = \frac{U}{2(W-a)}$$

Where $U$ is the area under the load-displacement curve in SENB test, and $W$ and $a$ are the width and initial crack length of the SENB sample respectively.